\documentclass{ws-p8-50x6-00}

\def\be{\begin{equation}}
\def\ee{\end{equation}}
\def\bea{\begin{eqnarray}}
\def\eea{\end{eqnarray}}
\def\ii{\rm{i}}
\def\tr{\rm{Tr}}
\def\nn{\nonumber}
\def\ex{\rm{e}}
\def\lsi{\raise0.3ex\hbox{$<$\kern-0.75em\raise-1.1ex\hbox{$\sim$}}}
\def\gsi{\raise0.3ex\hbox{$>$\kern-0.75em\raise-1.1ex\hbox{$\sim$}}}
\newcommand{\lsim}{\mathop{\lsi}}
\newcommand{\gsim}{\mathop{\gsi}}
\def\bfx{{\bf x}}

\begin{document}

\title{Debye screening in the QCD plasma}

\author{O. PHILIPSEN}

\address{CERN Theory Division, 1211 Geneva 23, Switzerland
\thanks{Address since October 2000: \em
Center for Theoretical Physics, Massachusetts Institute of Technology,
Cambridge, MA 02139, USA
} 
}


\maketitle

\abstracts{
Various definitions for the QCD Debye mass and its 
evaluation are reviewed in a non-perturabtive framework for the study of screening of
general static sources. While it is possible to perturbatively integrate over
scales $\sim T$ and thus construct a 3d effective theory, 
the softer scales $\sim gT$ and $\sim g^2T$ are strongly coupled
for temperatures $\lsim 10^7$ GeV and require lattice simulations.
Within the effective theory, a lattice treatment of screening at finite quark densities
$\mu \lsim 4/T$ is also possible.
}

\section{Introduction}

One goal of heavy ion collision experiments
is to probe the deconfined phase of QCD, in which 
matter is expected to exist in a state of 
quark-gluon plasma. 
At present, we are still far from a satisfactory understanding of its
properties, and hence of unambiguous signals for its detection.
Qualitative expectations
are based on asymptotic freedom, i.e.~the coupling weakening with increasing temperature.
However, for any reasonable temperatures below $10^{11}$ GeV, perturbative
treatments are badly convergent due to infrared-sensitive contributions. 
On the other hand, the only non-perturbative first principles approach, four-dimensional
(4d) lattice simulations, is severely restricted in the presence of light fermions, and
as of yet fails completely at finite baryon density.

Considerable progress has been made over the last years in studying the screening of
static sources in an equilibrated plasma by simulating dimensionally reduced QCD, as
well as 4d pure gauge theory.
In this contribution these developments are reviewed with special emphasis
on Debye screening, which is believed to play
a central role for deconfinement.
While at present it is not clear
whether screening of static charges can be directly probed experimentally,
it is a well-defined physical plasma property, providing the
dynamical length scales over which interactions are effective.

One difficulty with Debye screening in QCD is properly defining it.
After briefly recalling the QED analogue and pointing
out the differences in QCD, a non-perturbative framework to study screening
of a general static source is given. 
Next, an effective theory description for hot QCD
is reviewed, permitting a convenient non-perturbative study and 
interpretation of screening lenghts,
%pertaining to different gauge-invariant sources. 
before returning to the question of a gauge-invariant definition and non-perturbative evaluation 
of Debye screening. Finally, Debye screening in the presence of a finite baryon density
is discussed.

\subsection{Debye screening in QED}

Consider first a QED plasma in equilibrium, into which an external 
static charge is introduced. The modified Hamiltonian is then
\be
H'=H+H_{ext}=\frac{1}{2}\left((\bf{E}+\bf{E}_{ext})^2+\bf{B}^2\right)\;.
\ee
In linear response theory the perturbation of the electric field is
\cite{kap}
\bea
\delta\langle E_i(\bf{x})\rangle&=&\lim_{t\rightarrow\infty}\ii\int_{t_0}^t dt'\,
\tr\left(\rho [H_{ext}({\bf x}',t'),E_i(\bfx,t)]\right)\nn\\
&=&\lim_{t\rightarrow\infty}-\ii\int_{t_0}^t dt'\,\int d^3x'\,E_j^{ext}(\bf{x}') 
\langle[E_i({\bf x},t),E_j({\bf x}',t')]\rangle\,.
\eea
With $E_i(\bfx)=-\nabla_i V(r)$, the potential of the 
static source is determined by the correlation of electric fields, or 
equivalently by the Fourier
transform of the $A_0$ propagator, 
\be
V(r)=Q\int\frac{d^3k}{(2\pi)^3}\frac{\ex^{\ii kr}}{k^2+\Pi_{00}(0,\bf{k})}
=Q\frac{\ex^{-m_Dr}}{4\pi r}\,.
\ee
Thus, the electric field of the source is screened by the charges in the plasma, and
the Debye mass, or inverse screening length, is defined by the pole
of the photon propagator, 
\be
\Pi_{00}(k_00,{\bf k}^2=-m_D^2)=m_D^2.
\ee
The leading order one-loop diagram gives 
$m_D^0=eT/\sqrt{3} +{\cal O}(e^2 T)$.
On the other hand, there is no magnetic screening to all orders of perturbation theory, i.e.~
$\Pi_{ii}(k_0=0,{\bf k}\rightarrow 0)=0$.

\subsection{Non-abelian Debye screening}

Following the same steps in an SU(N) theory with $N_f$ fermions and
evaluating the pole of the one-loop $A_0$ propagator, one
finds the leading order result
\be
V(r)\sim\frac{\ex^{-m_D^0r}}{4\pi r}\,,
\quad m_D^0=\left(\frac{N}{3}+\frac{N_f}{6}\right)^{1/2} gT\,.
\ee
However, at next-to-leading order the problem becomes non-perturbative.
The general form of the series in $g$ can be shown to be
\cite{reb}
\be \label{dm}
m_D=m_D^0+{N\over4\pi}g^2 T\ln{m_D^0\over g^2T}
+c_{N}g^2T + {\cal O}(g^3T)\,,
\ee
which is non-analytic in the coupling constant. While
the coefficient of the logarithm is fixed perturbatively, 
$c_N$ is entirely non-perturbative. 
The reason is that, starting from this order in $g$, the
non-abelian $A_0$ couples to the soft magnetic gluons $A_i\sim g^2T$,
and hence becomes sensitive to the non-abelian infrared divergencies in the 
magnetic sector for which there is no perturbative cure \cite{linde}:
$\Pi_{ii}(k_0=0,{\bf k}\rightarrow 0)\sim g^2T\neq0$, with contributions
from all loop orders two and larger.

This raises the conceptual problem whether a perturbative definition 
of Debye screening in QCD is at all sensible. Moreover, neither $\Pi_{00}$ nor the
colour-electric field $\bf{E}$ are gauge-invariant physical concepts in a non-abelian
theory. It can be shown that the pole of the $A_0$ propagator is gauge-invariant order by order
in a (resummed) perturbation series \cite{kkr}. However, this does 
not guarantee the
existence of a pole in the full propagator. Even if it exists, 
its non-perturbative relation 
to physical quantities remains unclear in a situation where the potential
of a single static charge is not defined. 

There are then two ways to deal with this problem non-perturbatively.
One is to assume a
pole in the full propagator 
and to determine its value from the
exponential fall-off of the gauge-fixed propagator in space. Results of this 
approach are presented in another contribution to this conference \cite{petr}.
The second option, which will be discussed here, is to seek a manifestly gauge-invariant
definition of Debye screening.

\subsection{Non-perturbative description of screening}

Let us first recall the general
definitions needed to non-perturbatively describe screening of any gauge-invariant source.
Consider gauge-invariant, local operators $A(x)$. Static equilibrium physics is 
described by spatial correlation functions of euclidean 
time averages
\be \label{stateq}
C(|\bf{x}|)=\langle \bar{A}({\bf x})\bar{A}(0)\rangle_c\sim \ex^{-M|\bfx|},\quad
\bar{A}({\bf x})=T \int_0^{1/T} d\tau A(\bf{x},-\ii\tau)\,.
\ee
These fall off exponentially with distance. The 
``screening masses'' $M$ have a precise non-perturbative definition: they are the eigenvalues
of the spacewise tranfer matrix in the corresponding 
lattice field theory. Physically, they correspond to
the inverse length scale over which the equilibrated plasma is sensitive
to the insertion of a static source carrying the quantum numbers of $A$. Beyond
$1/M$, the source is screened and the plasma appears undisturbed.
With these definitions, all screening lengths corresponding to gauge-invariant sources
can be computed on the lattice in principle.

\section{Effective theory description}

If we are interested in length scales larger than the inverse temperature, $|\bfx|\sim 1/gT\gg
1/T$, the situation simplifies considerably. In this case the integration range 
for euclidean time averaging in Eq.~(\ref{stateq}) becomes very small, and the problem
effectively three-dimensional (3d). The calculation of the
correlation function $C(|\bf{x}|)$ can be factorized: 
the time averaging may be performed perturbatively
by expanding in powers of the scale ratio $gT/T\sim g$, 
which amounts to integrating out all modes with
momenta $\sim T$ and larger (i.e.~the non-zero Matsubara modes, in particular the fermions).
This procedure is known as dimensional reduction \cite{dr}.
The correlator $C(|\bf{x}|)$ of 3d fields is then to be evaluated
with a 3d purely bosonic effective action, describing the non-perturbative physics of the 
modes $\sim gT$ and softer, which can be done on the lattice. 
The screening masses $M$ now are the eigenvalues of the 3d tranfer matrix.
The 3d purely bosonic simulations are much easier and thus more accurate
than simulating directly in 4d. However, the perturbative integration over hard modes
and neglecting higher dimension operators in the effective action
introduces a relative error, which at two-loop level is \cite{rules}
\be
\frac{\delta C}{C}\sim {\cal{O}}(g^3)\,.
\ee
In the treatment of the electroweak phase transition, this error is less than
5\%\cite{ew}, but for hot QCD the coupling and thus the error is larger. An estimate
of its size will be given later, when results from the full and the effective theory 
are compared.

The effective theory emerging from hot QCD by dimensional reduction 
is the SU(3) adjoint Higgs model with the action
\begin{equation} \label{actc}
        S = \int d^{3}x \left\{ \frac{1}{2} \tr(F_{ij}F_{ij})
        +\tr(D_{i}A_0)^2 +(m_D^0)^2 \tr(A_0^2)
        +\lambda_3(\tr(A_0^2)^{2} \right\} .
\end{equation}
Apart from the SU(3) gauge symmetry, the action respects
SO(2) planar rotations, two-dimensional parity $P$,
charge conjugation $C$ and $A_0$-reflections $R$.
Its parameters are 
via dimensional reduction functions of the four-dimensional gauge
coupling $g^2$, the number of colours $N$ and flavours $N_f$, the fermion masses 
and the temperature $T$.
In all of the following fermion masses are assumed to be zero, 
but in principle any other values may be considered as well. 
At leading order in the reduction step one then has
\be \label{params}
g_3^2=g^2(\bar{\mu})T,\quad
\lambda_3=\frac{1}{24\pi^2}(6+N-N_f)g^4(\bar{\mu})T\,.
\ee
The reduction step has been performed to two-loop order \cite{ad} at which parameters 
have relative accuracy ${\cal O}(g^4)$.
Specifying $T$ and a renormalization scale $\bar{\mu}$ completely fixes 
the effective theory.

If we are interested in the longest scales of order
$\sim 1/g^2T$, one may be inclined to integrate out the fields $A_0 \sim gT$ as well,
leaving a 3d pure gauge theory. In order to decide which is the correct effective theory
in practice, 
simulation results for screening lengths will now be compared 
between the reduced and the full theory.

\begin{figure}[tb]%

\begin{center}
\leavevmode
\epsfig{file=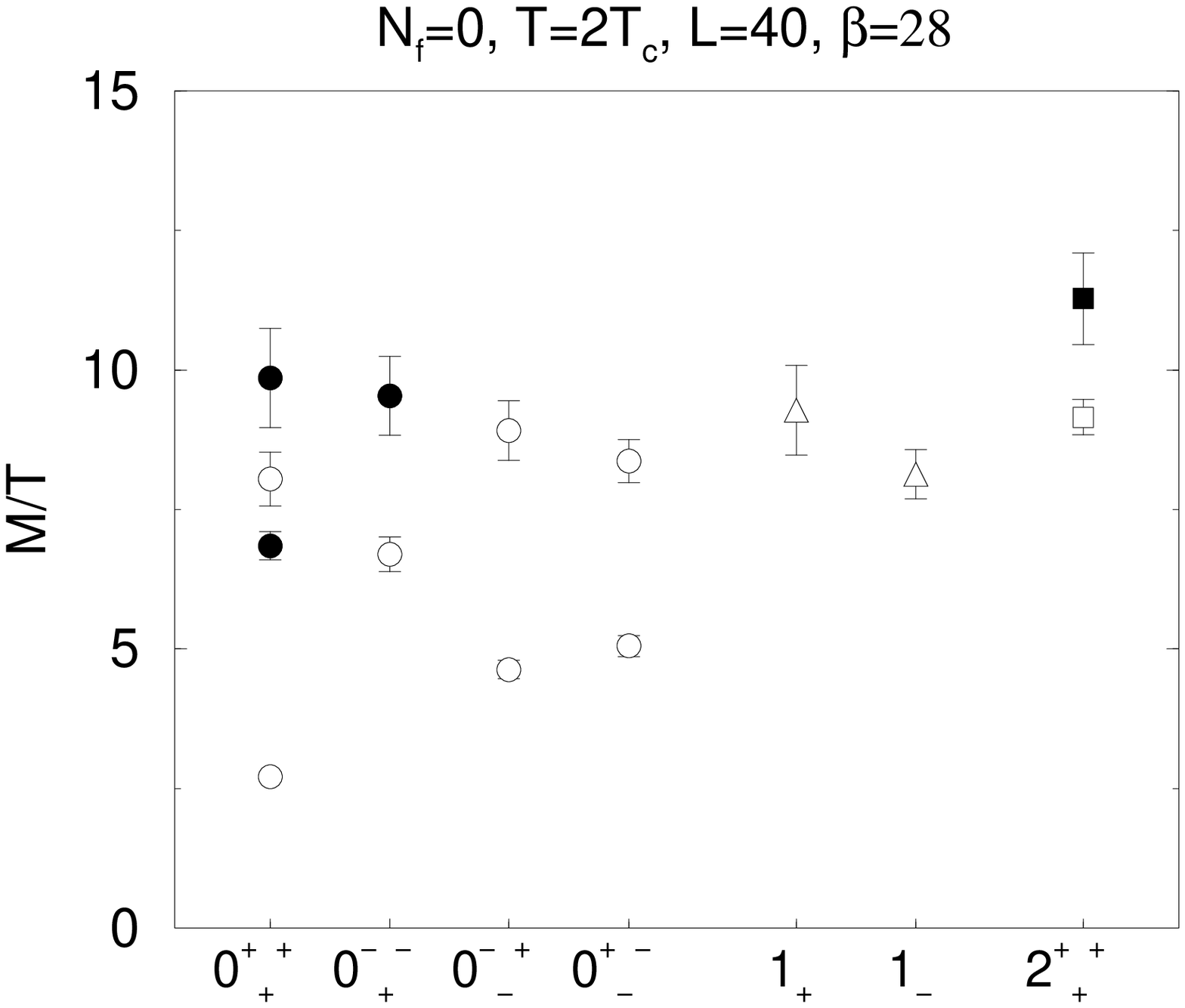,width=5cm}
\leavevmode
\epsfig{file=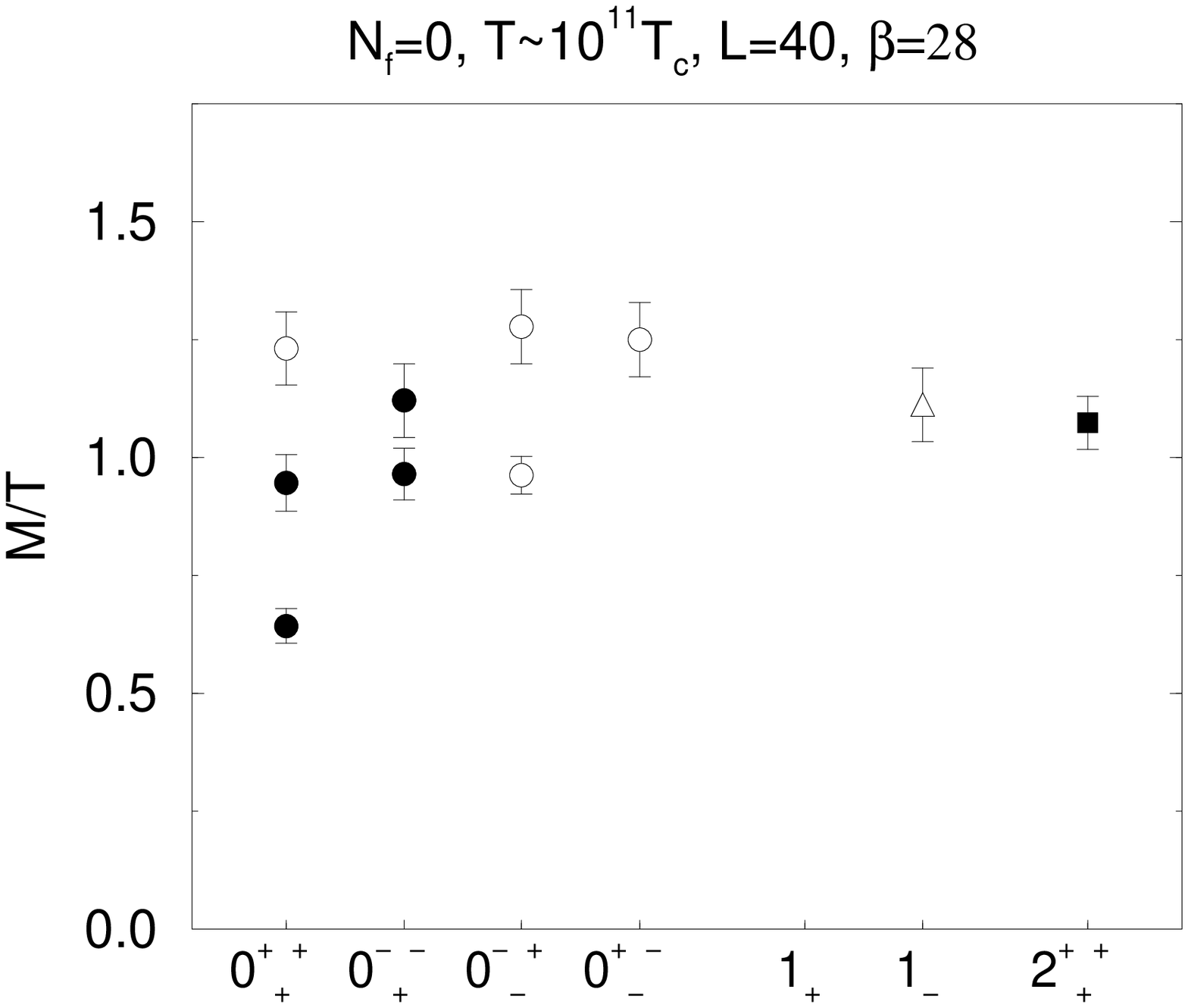,width=5cm}
\end{center}
\caption[]{\label{mt} Spectrum of screening masses in various quantum number channels
at low and high temperatures. Filled symbols denote 3d glueball states.
}
\end{figure}

Results from simulations \cite{hlp1} of the gauge-invariant screening spectrum with the 
effective theory Eq.~(\ref{actc}) are displayed in Fig.~\ref{mt}.
Screening masses are classified according to the symmetires of the 3d Hamiltonian by the
quantum numbers $J^{PC}_R$.
The open symbols show states dominated by scalar operators such as 
$\tr (A_0^2), \tr (A_0F_{12}^2)$ etc., whereas full symbols show states receiving 
only gluonic contributions $\tr (F_{12}^2)$ etc. It is a remarkable finding of 
detailed mixing analyses in several models \cite{mods,hp,hlp1}, 
that the latter are quantitatively
consistent with the 3d glueball states \cite{teper} and completely 
insensitive to the presence of the $A_0$ scalar field. Because of this pronounced 
non-mixing, we can state a first important result: For any reasonable temperatures
the largest correlation length of gauge-invariant, local  operators belongs to
the $A_0$ degrees of freedom and not to the $A_i$, in contrast to the naive parametric
picture. This demonstrates that the physics from the scale $gT$ down to $g^2T$ is
completely non-perturbative, and hence $A_0$ may not be integrated out perturbatively.
Only for temperatures $T\gsim 10^7T_c$ becomes the coupling small enough that the
parametrically expected ordering of screening masses is realized.

\begin{figure}[tb]%
\begin{center}
\leavevmode
\hspace*{-1in}
\epsfig{file=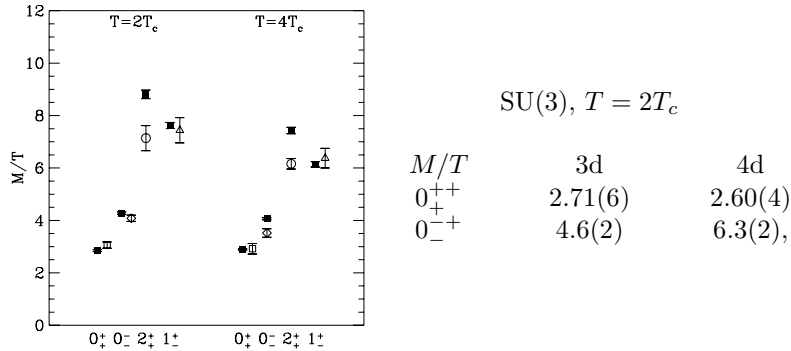,width=5cm}
\leavevmode
\begin{minipage}{1in}
\vspace*{-2in}
\begin{tabular}{ccc}
%\hline
& SU(3), $T=2T_c$& \\
&&\\
$M/T$      &  3d  & 4d\\
$0^{++}_+$ &  2.71(6)        & 2.60(4)    \\
$0^{-+}_-$ &  4.6(2)         & 6.3(2),     \\
%\hline
\end{tabular}
\end{minipage}
\end{center}
\caption[]{\label{spec} Comparison of screening masses in hot SU(2) (left) and SU(3) (right)
pure gauge theory as determined in 4d (empty symbols) \cite{dg} and 3d (full symbols) \cite{hp,hlp1} 
effective theory simulations.}
%\vspace*{-1cm}
\end{figure}

Next, we are interested in the accuracy of the reduced theory.
Fig.~\ref{spec} 
compares the results for hot SU(2) gauge theory as obtained
in the full lattice theory \cite{dg} with those from the effective theory \cite{hp}. Note that
the effective theory is only valid up to its cut-off $M/T\sim 2\pi$ and above this
level disagreement is to be expected. 
The apparent excellent agreement for the two lowest lying states should be taken
with care given some systematic uncertainties. In 4d simulations,
the projection properties of operators and the scaling behaviour 
are more difficult to control than in 3d. In the effective theory, there is an ambiguity
in the choice of renormalization scale. The uncertainty in the comparison may thus be
as large as 20\%. 
For the case of SU(3), one finds 
again quantitative agreement in the largest correlation length but about 20\%
deviation in the next shorter one.

One advantage of the effective theory approach is its easy inclusion
of fermions, which are treated analytically and just amount to changing 
the parameters Eq.~(\ref{params}) of the 
action to be simulated. However, the screening masses $M/T$ are found to increase
with $N_f$ \cite{hlp1}, and at $N_f=4$ the lowest of them is 
already of the order $\sim 1/2\pi T$ and hence not expected to be quantitatively accurate. 
Moreover, close to $T_c$ the
fermionic modes in the 4d theory feel the chiral phase transition, such that for
$T\lsim 2-3T_c$ the correlation length obtained
from pion operators becomes the longest one\cite{gg}. In this situation fermions may not
be integrated out and the purely bosonic effective theory is invalid.
We may thus conclude that dimensional reduction gives a reasonable description 
of the largest correlation lengths of hot QCD for temperatures 
$\gsim 2-3T_c$, with an error $\lsim 20\%$ that decreases with temperature.

\section{The Polyakov line and the heavy quark potential}

\begin{figure}[tb]%
\begin{center}
\leavevmode
\epsfig{file=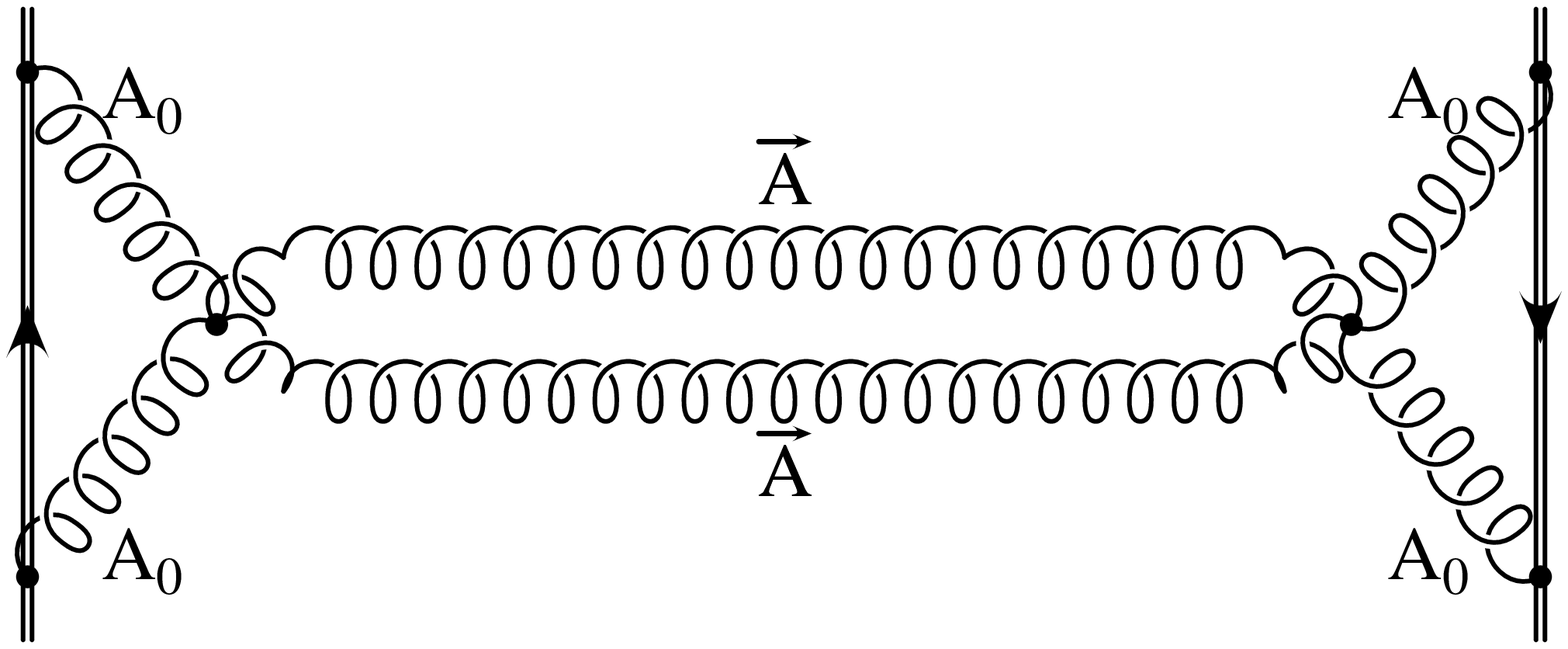,width=5cm}
\leavevmode
\hspace*{0.5cm}
\epsfig{file=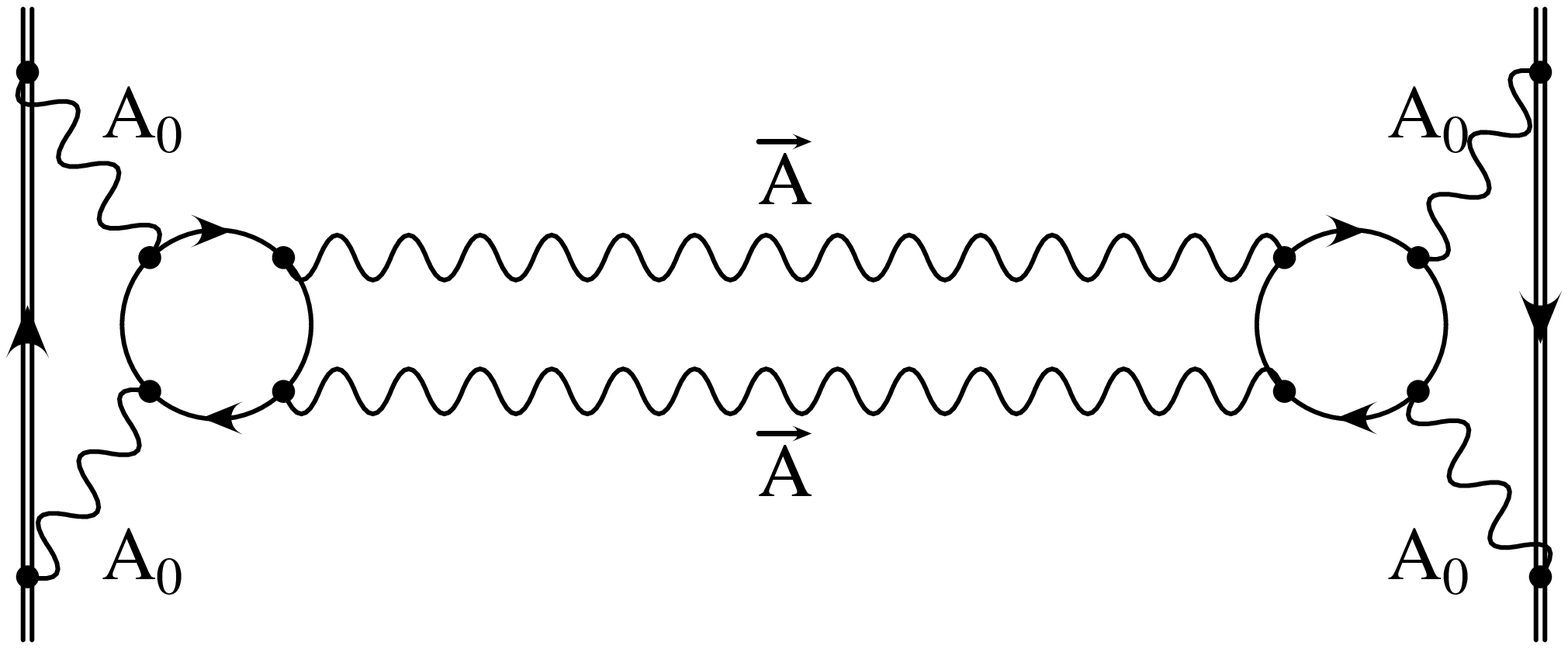,width=5cm}
\end{center}
\caption[]{\label{plhl} Higher loop diagrams contributing to the Polyakov line correlator in QCD (left)
and QED (right).}
\end{figure}

A gauge-invariant, non-local operator that has been suggested in the literature 
to describe Debye screening is the Polyakov line
\be
L({\bf x})=\frac{1}{N}\tr {\cal P} \exp -\ii g\int _0^\beta d\tau\, A_0(\bf{x},\tau)\;.
\ee
Its spatial correlator is related to the free energy of a quark anti-quark
pair in the plasma, and thus to the static potential at finite temperature \cite{ms}
\be
\frac{\langle L^\dag(r)L(0)\rangle}{\langle|L|\rangle^2}=\exp -\beta F_{q\bar{q}}=\exp-\frac{V(r)}{T}\,.
\ee
The idea is that by considering a source together with an anti-source gives
a manifestly gauge-invariant operator related to heavy quark systems. 
In leading order perturbation theory one has
$L=1-g^2/2\tr(\int d\tau A_0)^2+\ldots$, and hence the correlator at this order 
is given by the exchange of two electric gluons leading to
\be
\langle L^\dag(r)L(0)\rangle_c =\frac{(N^2-1)g^4}{8N^2T^2}\frac{\ex^{-2m_D^0r}}{(4\pi r)^2}\;,
\ee
with the leading order Debye mass $m_D^0$. However, this state of affairs is spoilt 
by higher loop diagrams,
as several authors have pointed out \cite{plit,ay}. 
Fig.~\ref{plhl} left shows a diagram of 
order $g^8$, where now $A_0$ can couple to two magnetic gauge fields $A_i$, and hence
the exponential fall-off of this diagram is governed purely by magnetic gauge fields.
Indeed, through a fermion loop a similar coupling is possible even in QED, cf.~Fig.~\ref{plhl}
right.
There, magnetic gauge fields are strictly massless, and the 
correlator falls off algebraically rather than exponentially. It thus
fails to produce the Debye mass for electric fields even in QED.

With our previous discussion of screening masses at hand, it is not difficult
to understand how the Polyakov line correlator should fall off away from $T_c$.
%\footnote{
%The argument is not applicable in SU(3) Yang-Mills for $T\sim T_c$, where
%the Polyakov line is sensitive to tunnelling between $Z_N$ broken and symmetric phases.}
$L$ contains a sum over all $J^{PC}$ sectors in the fundamental representation.
Furthermore, because of gauge-invariance it couples only to gauge-invariant 
intermediate states. Its correlator should therefore asymptotically fall off
with the lightest gauge-invariant screening mass of the spectrum.
Indeed, a recent 4d simulation of the SU(3) Polyakov line correlator \cite{kkll}
finds it exponential decay to be governed by $M/T\approx 2.5(3)$, which is fully compatible
with the lightest $0^{++}_+$ mass given in the table in Fig.~\ref{spec}. The same observation
is made to rather high precision in a hot (2+1)-dimensional gauge theory \cite{bp}.

\section{A gauge-invariant definition for $m_D$}

A few years ago it was suggested to 
use a symmetry to distinguish between the electric and the magnetic sectors
of the theory \cite{ay}, namely euclidean time reflection ${\cal T}: \tau\rightarrow -\tau$.
This operation changes the sign of the electric gauge fields, $A_0\rightarrow -A_0$,
but not that of the magnetic ones. 
The non-perturabtive Debye screening length $1/m_D$ may then be defined to be the largest
correlation length measured from a correlator
$\langle A(\bf{x}) B(\bf{0})\rangle$, where $A,B$ are gauge-invariant operators
which are local in 3-space and odd under ${\cal T}$.
Examples of four-dimensional operators
with these properties are $\tr ({\rm Im} L), \tr({\rm Im}L F_{12}), \tr(F_{03}F_{12}^2),
\tr(F_{03}F_{12})$ etc. 

In the 3d effective theory, where "time" has been
reduced away, the remnant of euclidean 
time reflection is the scalar reflection symmetry  $R: A_0\rightarrow -A_0$. 
The Debye mass then corresponds to the lightest eigenstate 
with quantum number $R=-$. 
It is now easy to go back to the results for the spectrum in Fig.~\ref{mt}
and identify the lowest mass in the $0^{-+}_-$ channel as the Debye mass according to this 
definition\footnote{
Alternatively, it was proposed to measure
matrix elements between $R$-odd operators and a $Z_N$ domain wall, introduced
by a twist in the effective action \cite{bk}.}.
The corresponding operator 
%whose correlation produces this mass in the
%3d effective theory 
is $\tr (A_0 F_{ij})$, which may be interpreted as an electric
gauge field dressed by magnetic gluons to make it gauge-invariant. 
Despite its coupling to $A_i$, 
$R$-symmetry prohibits its correlation to be governed by the 3d pure gauge sector.
This definition certainly
refers to a physical quantity. 
However, as it couples to colour singlet sources, its relation
to screening of colour flux and to deconfinement of a quarkonium system is not obvious.

\subsection{Perturbative and non-perturbative contributions to $m_D$}

We may now ask to what extent this Debye mass can be computed in
perturbation theory. First, note that the perturbative expansion follows the same pattern as
in Eq.~(\ref{dm}). At tree level, the correlator is given by exchange of one electric and two
magnetic gluons. Since the latter are massless at tree level, the leading order
fall-off is yet again $m_D^0$. However, at higher orders
the contributions are different from those to the $A_0$ propagator.

There is a recipe to compute the coefficient $c_N$ non-perturbatively \cite{ay}:
if temperature is sent to infinity,
the $A_0$ becomes infinitely heavy and decouples,
so it may be integrated out.
In this case, $A_0$ turns into a static external source, and the only dynamical fields
left are the $A_i$ of a 3d pure gauge theory. We thus have
\be
\langle \tr(A_0F_{ij})(\bfx)\tr(A_0F_{lm})(0)\rangle
\longrightarrow \frac{\ex^{-m_D^0|\bfx|}}{4\pi |\bfx|}
\langle F^a_{ij}(\bfx)W_{ab}(\bfx,0)F_{lm}^b(0)\rangle \;,
\ee
where $W_{ab}(\bf{x},0)$ is an adjoint representation Wilson line.

As a result, our correlator factorizes into the free $A_0$ propagator
and a correlator of the magnetic field strength, connected by an adjoint Wilson
line representing the static $A_0$. The field strength correlator falls 
off exponentially
$\sim \exp-(\Delta+c_N)|\bfx|$, where $\Delta$ contains a logarithmic divergence
which gets absorbed by the renormalisation of the $A_0$ self-energy \cite{ay}.

Measuring $c_N$ in a 3d pure gauge theory \cite{lp1} and comparing with the measurements
of the full $m_D$ we can now summarise the various contributions to the SU(3) Debye
mass in Table \ref{debc}.
\begin{table}
\caption[]{\label{debc}
The different contributions to the Debye mass, Eq.~(\ref{dm}). (1+2) refers to the 
first two terms in units of $g^2T$.}
\begin{center}
\begin{tabular}{|c|r@{.}l|r@{.}l|r@{.}l|r@{.}l|}
\hline
 &\multicolumn{2}{|c|}{$m_D/g^2T$} &
\multicolumn{2}{|c|}{$(1+2)$} &
\multicolumn{2}{|c|}{$c_3$} &
\multicolumn{2}{|c|}{${\cal O}(g^3T)/(g^2T)$} \\
\hline
$T=2T_c$  & 1&82 (3) & 0&514 & 1&65(6) & -0&35(8) \\
$T=10^{11}T_c$  & 3&83 (9) & 2&165 & 1&65(6) & 0&015(90) \\
\hline
\end{tabular}
\end{center}
\end{table}
We observe that at reasonable temperatures the next-to-leading order coefficient
$c_3$ is larger than the leading contribution, and thus the Debye mass is entirely
non-perturbative. Again the reason is its coupling
to the non-perturbative magnetic sector $\sim g^2 T$. 
Note also that the ${\cal O}(g^3T)$ contributions are small
compared to the $g^2T$ contributions. Together with the previous observations about scales involved,
one may then conclude that the physics of
screening for all temperatures of pratical interest is dominated by the scale $\sim g^2T$.
Only at asymtotically high temperatures $T\gsim 10^7 T_c$ is the perturbative picture restored. 

\section{Debye screening at finite baryon density}

In heavy ion collisions the initial state has non-vanishing
baryon number, and so does any subsequent plasma state.
We would therefore like to know if and how a chemical potential for 
quarks affects the physics of screening. Unfortunately, finite 
baryon density cannot be addressed
by standard lattice QCD because of the so-called ``sign-problem'': when a chemical
potential term is added to the theory, the fermion determinant becomes
complex prohibiting Monte Carlo importance sampling, which requires a strictly positive measure. 
To date, no working
cure has been found for this problem \cite{alf}.

Recently it was demonstrated that in dimensionally reduced finite density
QCD the sign problem is in fact numerically tractable \cite{hlp2}.
Inclusion of a chemical potential term for quarks leads
to one extra term in the action Eq.~(\ref{actc}) and changes the tree level Debye mass,
\be
S \to S + \ii \frac{\mu}{T}\frac{N_f}{3\pi^2}\int d^3x\,\tr A_0^3 \,,\quad
m_D^0\to m_D^0\biggl[1+\Bigl( \frac{\mu}{\pi T}\Bigr)^2 \frac{3 N_f}{2 N + N_f}
\biggr]\,.
\ee
This action is complex and cannot be simulated
with standard methods. Instead, a reweighting procedure has to be performed, where
the complex term of the action is absorbed into the observable, so that
\be
\langle {\cal O}\rangle = 
\frac{\langle {\cal O} \ex^{-\ii S_\mu}\rangle_0}{\cos(S_\mu)\rangle_0}\,,
\ee 
and the averaging is done with $\mu=0$.
The sign problem, i.e.~cancelling contributions to the expectation value, occur whenever
$S_\mu\gg 1$. Fortunately, the width of the distribution of this quantity, growing
as $\sim (\mu/T) V^{1/2}$, is narrow enough to permit
large enough volumes for simulated masses to reach their infinite volume levels,
as long as $\mu\lsim 4T$ \cite{hlp1}.
\begin{figure}[tb]%
\begin{center}
\leavevmode
\epsfig{file= 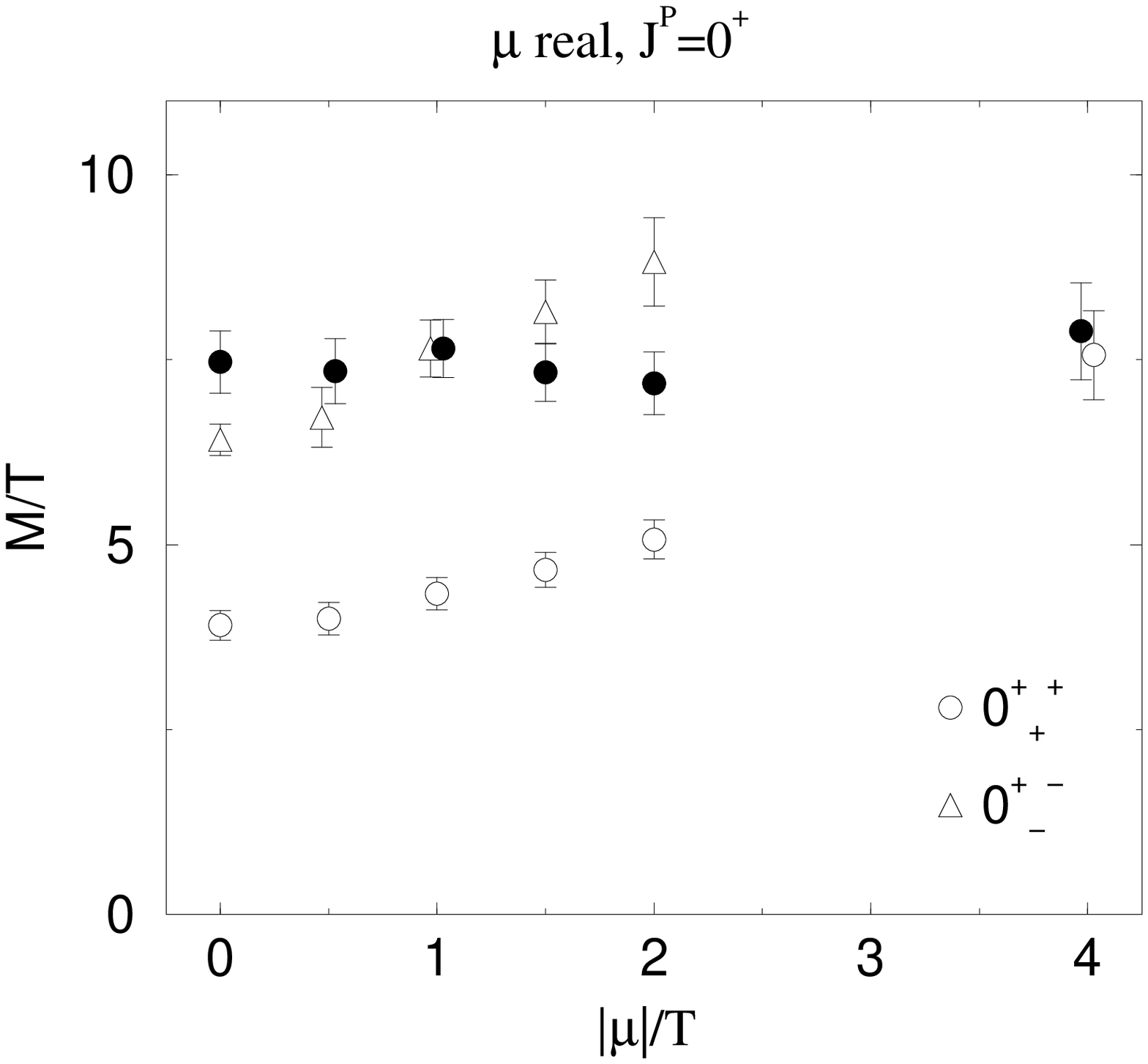,width=5cm}
\leavevmode
\hspace*{0.5cm}
\epsfig{file=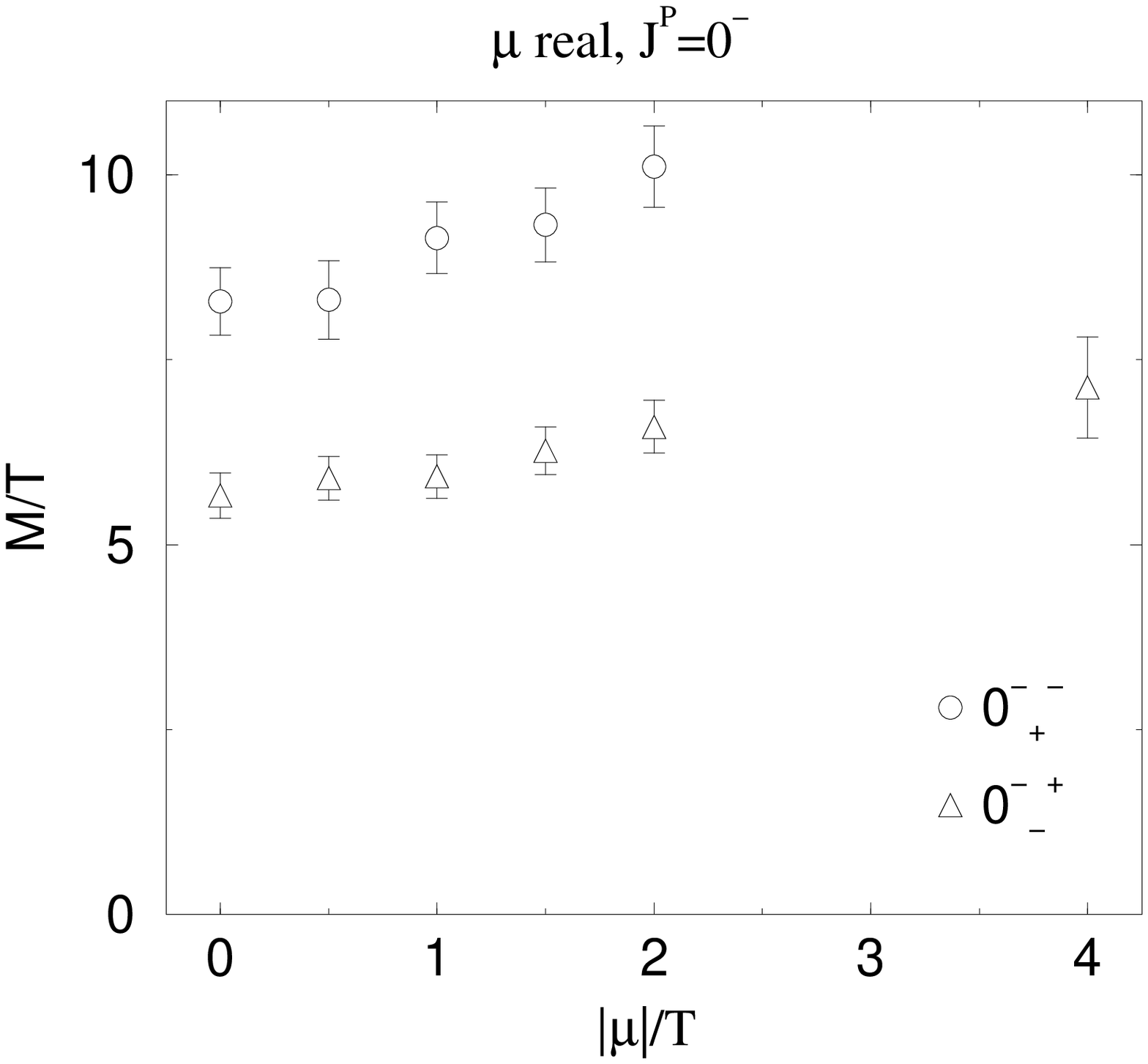,width=5cm}
\end{center}
\caption[]{\label{mmu}
The behaviour of the lowest screening masses when a chemical potential is switched on.
The triangles in the right panel give the Debye mass according to the gauge invariant definition
in the last section.
}
\end{figure}

There is one important change of the symmetries
of the theory, which is relevant for Debye screening \cite{ay}: 
the chemical potential term violates $R$ and $C$, and thus only $P$ is
left as a good symmetry to classify the screening states. Since the definition of $m_D$ was
based on the use of $R$, it seems no longer applicable in the finite density situation.
%One may argue \cite{ay} that, in the case of little mixing,
%the pole in the corresponding correlator will be slightly displaced from
%$p=im_D$ off the imaginary axis to $p=im_D + \Delta$. 
In the 3d effective theory, the corresponding
state can mix with any lighter $R=+$ states and decay. 
However, for $\mu=0$, we found the Debye mass
to be the lowest state in the $0^{-+}_-$
channel. 
This is also the lowest state with $P=-$.
%For an infinitesimal $\mu$ switched on, it thus remains the
%lightest state in the $0^-$ channel, which cannot decay.
As Fig.~\ref{mmu} shows, this remains to be the case as 
$\mu/T$ is switched on,
and by continuity one may now identify the lowest $0^-$ mass to correspond
to the Debye mass. 

Finally, it has been demonstrated, that screening masses for $\mu\lsim T$ should be 
accessible to 4d simulations with imaginary 
$\mu$ and analytic continuation\cite{hlp2}.

\section{Conclusions}

The analogy of
Debye screening in QED and QCD is rather limited, because in QCD it
is not possible to define the electric field induced by an isolated charge: 
neither of these are gauge-invariant physical concepts. The problem then is
to specify what we mean by Debye screening in QCD. 
Depending on the definition, one gets different answers and different interpretations of the results.
Defining the Debye mass through the pole of 
the full $A_0$ propagator has the disadvantage that its existence is
not guaranteed, and its non-perturbative connection to physical quantities is not clear. Employing
the Polyakov line correlator describing a heavy quark anti-quark pair
does not probe Debye screening: it mixes electric and magnetic sectors, failing to 
fall off with $m_D$ even in QED.
A gauge-invariant definition is possible through correlations of operators odd
under euclidean time reflection, which reflects $A_0$ but not $A_i$.
 
In general, the screening of static sources in equilibrium is a well-defined problem
that has been studied non-perturbatively by lattice QCD and with effective field theory methods.
A detailed picture of the static length scales for non-abelian plasma physics 
has emerged. An important lesson for constructing effective theories is that, 
for all temperatures $T<10^7$ GeV, the scales $\sim gT$ and $\sim g^2T$ 
are not separable non-perturbatively. All static screening physics seems to be dominated
by the dynamics on soft scales $\sim g^2T$.

Finally, dimensionally reduced QCD gives a realistic description of static
phenomena down to temperatures of a few $T_c$, and is able to non-perturbatively
accomodate chemical potentials for quarks $\mu\lsim 4T$. It can therefore address the physics of
screening in the parameter range relevant for heavy ion collisions.

\end{document}